\documentclass[12pt]{article}

\usepackage{longtable}
\usepackage{amsmath}
\usepackage{amssymb}
\usepackage{algorithm}
\usepackage{latexsym}
\usepackage{epsfig}
\usepackage{epsf}
\usepackage{fullpage}
\usepackage{setspace}
\usepackage{multirow}
\usepackage{dcolumn}
\usepackage{url}

\include{math_operators}

\newcounter{xxx}
\def \ve{\mbox{\boldmath $e$ \unboldmath}\!\!}
\def \vZ{\mbox{\boldmath $Z$ \unboldmath}\!\!}

\def \vbeta{\mbox{\boldmath $\beta$ \unboldmath}\!\!}

\def \valpha{\mbox{\boldmath $\alpha$ \unboldmath}\!\!}

\def \vZ{\mbox{\boldmath $Z$ \unboldmath}\!\!}
\def \vA{\mbox{\boldmath $A$ \unboldmath}\!\!}

\def \vx{\mbox{\boldmath $x$ \unboldmath}\!\!}

\def \vv{\mbox{\boldmath $v$ \unboldmath}\!\!}

\def \vy{\mbox{\boldmath $y$ \unboldmath}\!\!}

\def \vtheta{\mbox{\boldmath $\theta$ \unboldmath}\!\!}

\begin{document}
%\firstpage{1}

\author{Yunlong Nie$^{2}$, Eugene Opoku$^{1}$, Laila Yasmin$^{1}$, Yin Song$^{1}$\\
Jie Wang$^{2}$, Sidi Wu$^{2}$, Vanessa Scarapicchia$^{3}$, Jodie Gawryluk$^{3}$\\ 
Liangliang Wang$^{2}$, Jiguo Cao$^{2}$, Farouk S. Nathoo$^{1,*}$\\\\
$^{1}$Department of Mathematics and Statistics, University of Victoria\\
$^{2}$Statistics and Actuarial Science, Simon Fraser University\\
$^{3}$Department of Psychology, University of Victoria\\
$^{*}$nathoo@uvic.ca\\}

\title{Spectral Dynamic Causal Modelling of Resting-State fMRI: Relating Effective Brain Connectivity in the Default Mode Network to Genetics}
%\author[Sample \textit{et~al}]{Keelin Greenlaw$^{1}$, Elena Szefer\,$^{2}$, Jinko Graham\,$^{2}$, Mary Lesperance$^{1}$, and Farouk S. Nathoo\,$^{1,}$\footnote{to whom correspondence should be addressed}; For the Alzheimer's Disease Neuroimaging Initiative}
%\address{$^{1}$Mathematics and Statistics, University of Victoria, Victoria, British Columbia, PO BOX 1700 STN CSC, Canada.\\
%$^{2}$Statistics and Actuarial Science, Simon Fraser University, Burnaby, British Columbia, V5A 1S6, Canada.}

\maketitle

\begin{abstract}
We conduct an imaging genetics study to explore how effective brain connectivity in the default mode network (DMN) may be related to genetics within the context of Alzheimer's disease and mild cognitive impairment. We develop an analysis of longitudinal resting-state functional magnetic resonance imaging (rs-fMRI) and genetic data obtained from a sample of 111 subjects with a total of 319 rs-fMRI scans from the Alzheimer's Disease Neuroimaging Initiative (ADNI) database. A Dynamic Causal Model (DCM) is fit to the rs-fMRI scans to estimate effective brain connectivity within the DMN and related to a set of single nucleotide polymorphisms (SNPs) contained in an empirical disease-constrained set which is obtained out-of-sample from 663 ADNI subjects having only genome-wide data. 
  
We examine longitudinal data in both a 4-region and an 6-region network and relate longitudinal effective brain connectivity networks estimated using spectral DCM to SNPs using both linear mixed effect (LME) models as well as function-on-scalar regression (FSR). In the former case we implement a parametric bootstrap for testing SNP coefficients and make comparisons with p-values obtained from the chi-squared null distribution. We also implement a parametric bootstrap approach for testing regression functions in FSR and we make comparisons between p-values obtained from the parametric bootstrap to p-values obtained using the F-distribution with degrees-of-freedom based on Satterthwaite's approximation. 

In both networks we report on exploratory patterns of associations with relatively high ranks that exhibit stability to the differing assumptions made by both FSR and LME.

\end{abstract}

\section{Introduction}\label{intro}

Alzheimer's Disease (AD) is a neurodegenerative disorder characterized by cognitive decline and progressive dementia and is thought to be caused by aberrant connections between cerebral regions involved in cognitive functioning (Li et al., 2013). Imaging genetics is an important area of scientific investigation in the search for genetic biomarkers of neurodegenerative disease, and in increasing our understanding of the genetic basis of brain structure and function in health and disease. 

The development of analytical methods for the joint analysis of neuroimaging phenotypes and genetic data is an important area of statistical research with many challenges. Recent reviews are provided in Liu and Calhoun (2014) and Nathoo et al. (2019). A great deal of work in imaging genetics has focussed on methods and analysis for examining the relationship between brain structure and genetics (see e.g., Stein et al., 2010; Hibar et al., 2011; Ge et al., 2012; Zhu et al., 2014; Greenlaw et al., 2017; Szefer et al., 2017; Lu et al., 2017; Song et al., 2019). Thompson et al. (2013) give an extensive overview of methods for the analysis of genetic data and brain connectivity with a broad focus on both diffusion tensor imaging (DTI) and fMRI data. These authors discuss the heritability of both structural and functional brain connectivity. Methods for the analysis of brain connectivity with an emphasis on structural connectomes is discussed  in Zhang et al. (2018, 2019).

Our focus in this article is in exploring potential associations between brain connectivity and genetics within the context of Alzheimer's disease and mild cognitive impairment. Effective brain connectivity and causal inference is discussed in Lindquist and Sobel (2016), and functional connectivity analysis for fMRI data is reviewed in Cribben and Fiecas (2016). Patel et al. (2006) and Chen et al. (2016) develop Bayesian approaches for modelling brain connectivity and Bowman et al. (2012) consider the analysis of fMRI functional connectivity using a multimodal approach. Here, effective connectivity refers to a directed measure of dependence from one brain region to another (see, e.g., Friston, 1994), while functional connectivity refers to the correlation between measured time series over different locations. 

In this paper we conduct an analysis examining the relationship between genetics and effective brain connectivity as measured by rs-fMRI within the default mode network (DMN).  The DMN consists of a set of brain regions that tend to be active in resting-state, when a subject is mind wandering with no intended task. In this state DMN regions will exhibit low frequency signals that tend to couple together. We consider networks comprised of four (DMN4 - 16 connections) and subsequently six (DMN6 - 36 connections) core regions of the DMN. The DMN4 analyses are based on network nodes located at the medial prefrontal cortex (MPFC), the posterior cingulate cortex (PCC), the left and right intraparietal cortex (LIPC and RIPC) with the Montreal Neurological Institute (MNI) locations for these regions depicted in Figure 1. The DMN6 analyses are based on six regions whose MNI locations are specified in Table 1. The regions we consider in DMN4 are a subset of the regions considered in DMN6 which, in addition to the original four regions, have added the left inferior temporal region (LIT) and right inferior temporal region (RIT). 

Our analyses involve examining effective connectivity networks from rs-fMRI data using Dynamic Causal Modeling (DCM; Li et al., 2011; Friston et al., 2003; Friston et al., 2014;  Razi et al., 2015; Friston et al., 2017), a nonlinear state-space framework for inferring interaction between latent neuronal states. We apply DCM to rs-fMRI using the SPM12 (v7219) software (Penny et al., 2011). Resting-state fMRI data are examined with the goal of investigating the potential interaction between different areas of the brain and to explore the potential association between this neuronal interaction in four (DMN4) and then eight (DMN6) core regions of the DMN and SNPs that are contained in a disease-constrained set. The DCM framework leads to directed networks and these networks are related to genetic data using longitudinal rs-fMRI analysis. Previous literature focussing on AD has found alterations to both effective and functional resting-state connectivity in the DMN (see e.g., Wu et al., 2011; Yan et al., 2013; Luo et al., 2018). Zhong et al. (2014) conduct an rs-fMRI study and demonstrate changes in directed functional connectivity in the DMN for subjects with AD, while Dipasquale et al. (2015) apply high-dimensional independent component analysis (ICA) to an rs-fMRI study and demonstrate changes in functional connectivity in subjects with AD. 

Our specific choice of regions which represent the nodes of the network is motivated by existing literature examining connectivity in the DMN and establishing the heritability of effective connectivity for networks based on these regions. These regions are a subset of the DMN regions considered in the rs-fMRI study of Wu et al., (2011), which demonstrated altered DMN functional and effective connectivity in AD, and for DMN4 they are the same regions considered in the rs-fMRI study of Sharaev et al. (2016), which investigated internal DMN relationships. Xu et al. (2017) examine effective connectivity in these four regions of the DMN using DCM and structural equation modelling in a twin study based on a sample of $n = 46$ pairs of twins with rs-fMRI. These authors find evidence for the heritability of effective connectivity in this network. We note that this study uses stochastic DCM rather than spectral DCM and the two approaches are not equivalent. They estimate the heritability of DMN effective connectivity in these regions to be 0.54 (that is the proportion of variability in DMN effective connectivity that can be attributed to genetics is 0.54). Their study provides evidence that there are genes involved in DMN effective connectivity for the network nodes depicted in Figure 1. This work paves the way for our study of relationships between effective connectivity and genetic markers in the same network. 

Glahn et al. (2010) use an extended pedigree design and rs-fMRI to examine genetic influence on functional connectivity within the DMN. Their study estimates the heritability of DMN functional connectivity to be 0.424 $\pm$ 0.17. That is to say, these authors estimate the proportion of variability of DMN functional connectivity that can be attributed to genetics to be 0.42 $\pm$ 0.17. This estimate is within error bounds of the heritability estimate of 0.54 obtained by Xu et al. (2017) for DMN effective connectivity. Importantly, Glahn et al. (2010) also suggest that the genetic factors that influence DMN functional connectivity and the genetic factors that influence gray matter density in these regions seem to be distinct. This result then motivates the search for genetic markers associated with DMN connectivity. Stingo et al. (2013) focus on relating brain connectivity to genetics and develop a Bayesian hierarchical mixture model for studies involving fMRI data. The mixture components of the proposed model correspond to the classification of the study subjects into subgroups, and the allocation of subjects to these mixture components is linked to genetic covariates with regression parameters assigned spike-and-slab priors.

Our strategy for data analysis proceeds as follows. We view both disease and rs-fMRI as measures of the brain and our fundamental interest is to relate these measures of the brain to genetics. The former is a coarse measure with three categories while the latter is a far a more detailed measure allowing for the investigation of statistical dependencies in the temporal rs-fMRI signal at different regions.  

We use an out-of-sample genome-wide scan of disease to select a priority subset of SNPs and this serves as a constraint on the SNPs that we relate to effective brain connectivity in subsequent analysis. We relate effective brain connectivity as characterized through spectral DCM to the empirical disease-constrained subset of genetic variables using longitudinal analyses based on both linear mixed effect (LME) models and function-on-scalar (FSR) regression (Morris, 2015). When testing for SNP effects in both models we compare p-values obtained from standard asymptotic distributions to those obtained from the parametric bootstrap.

%We do this first by looking at baseline rs-fMRI data from the ADNI2 study in four regions of the DMN (refer to figure and Table 1). Subsequently, we examine longitudinal rs-fMRI data in the same four regions leading to increased efficiency resulting from repeated measures. Finally, we expand the longitudinal analysis to an eight region network with 64 connections (for the regions depicted in Table 1) including hippocampal regions. Throughout, our examination of the data emphasizes detection of signals that appear consistently accross disease, baseline and longitudinal rs-fMRI analysis and across the application of different methods.

The remainder of the paper proceeds as follows. In Section 2 we discuss the neuroimaging and genetic data preprocessing, we present basic summaries and the use of spectral DCM for characterizing effective connectivity. The underlying state-space model and its estimation is discussed briefly. In Section 3 we examine how the genetic data are related to disease using a GWAS with 1,220,955 SNPs and 663 ADNI subjects not having rs-fMRI data. In doing so, we obtain an out-of-sample empirical disease-constrained set of SNPs for subsequent analysis examining the association between effective brain connectivity and genetics. In Section 4.1 we relate effective connectivity in DMN4 obtained from longitudinal rs-fMRI to genetics using LME and FSR respectively. In Section 4.2 we apply this same analysis to DMN6 and make comparisons between asymptotic null distributions and parametric bootstrap empirical null distributions. The paper concludes in Section 5 with a summary of our primary findings, a discussion of our analysis and its limitations, follow-up analyses and possible new methodological development that is motivated by the current work.

\section{Data and Preprocessing}\label{data}

Data used in the preparation of this article were obtained from the Alzheimer's Disease Neuroimaging Initiative (ADNI) database (adni.loni.usc.edu). The ADNI was launched in 2003 as a public-private partnership, led by Principal Investigator Michael W. Weiner, MD. The primary goal of ADNI has been to test whether serial magnetic resonance imaging (MRI), positron emission tomography (PET), other biological markers, and clinical and
neuropsychological assessment can be combined to measure the progression of mild cognitive impairment (MCI) and early AD. For up-to-date information, see \url{www.adni-info.org}. ADNI is an ongoing, longitudinal, multicenter study designed to develop clinical, imaging, genetic, and biochemical biomarkers for the early detection and tracking of AD. 

The selection criteria for our sample is as follows. We first begin with ADNI2 subjects (1437 at this stage) and consider those subjects with genome-wide data (774 left at this stage) and also with at least one resting-state fMRI scan at 3T (111). This leads to 111 subjects comprising  36 cognitively normal (CN), 63 MCI and 12 AD subjects, with these subjects having a mean age of 73.8 years with the range being 56.3-95.6 years, 46 of these subjects being male, 5 being left-handed, and with education measured in years ranging from 11 to 20.  

Table 2 presents several summary statistics associated with our sample, including a summary on the Apolipoprotein E (APOE) gene. The APOE gene is a known genetic determinant of AD risk and individuals carrying the $\epsilon$4 allele are at an increased risk of AD (see, e.g., Liu et al. 2013). Table 2 summarizes the number of APOE $\epsilon$4 alleles for the subjects in each disease category. In line with expectations from the literature (Genin et al., 2011), a signal from the APOE gene is present in the data (p-value = 0.0045; Fisher's Exact Test) with the AD group having a higher percentage of subjects with at least one $\epsilon$4 allele. In fact, the data summaries in Table 2 indicate that all but one of the AD subjects have at least one $\epsilon$4 allele of the APOE gene.

Diagnostic classification of AD participants was made by ADNI investigators according to diagnostic criteria for probable AD established by the National Institute of Neurological and Communicative Disorders and Stroke and the Alzheimer's Disease Related Disorders Association (NINCDS-ADRA; McKhann et al., 1984). Participants in the AD cohort also exhibited abnormal memory function on the Logical Memory II subscale of the revised Wechsler Memory Scale (WMS II, $\le$ 8 for 16 years education and above), a Mini Mental State Exam (MMSE) between 20 and 26 (inclusive), and a Clinical Dementia Rating of 0.5 (very mild) or 1 (mild). All control participants were free of memory complaints and deemed cognitively normal based on clinical assessments by the site physician showing an absence of significant impairment in cognitive functioning and performance of daily activities. Participants in the control cohort also exhibited normal memory function on the Logical Memory II subscale of the revised WMS (WMS II, $\le$ 9 for 16 years of education and above), a MMSE score between 24 and 30 (inclusive), and a Clinical Dementia Rating of 0.  

As described in Bondi et al. (2014), the ADNI criteria for MCI are: 1) subjective memory complaints reported by themselves, study partner, or clinician; 2) objective memory loss defined as scoring below an education-adjusted cut-off score on delayed recall of Story A of the WMS-R Logical Memory Test (score =8 for those with =16 years of education; score =4 for those with 8-15 years of education; score =2 for those with 0-7 years of education); 3) global CDR score of 0.5; and 4) general cognitive and functional performance sufficiently preserved such that a diagnosis of dementia could not be made by the site physician at the time of screening. %For more information on group classifications, including all additional eligibility criteria, please consult the ADNI2 procedures manual (ADNI, 2008).

Our analysis examines longitudinal rs-fMRI data from 111 subjects with repeated measurements resulting in a total of 424 rs-fMRI scans to begin with. Each scan leads to a network which has an associated follow-up time. As the longitudinal data in this study are quite sparse, we apply the Principal Analysis by Conditional Expectation (PACE; Muller, 2008) method which focuses on recovering the entire continuous-time temporal trajectory of the network edge parameters. As a result of the sparsity of the longitudinal data, it is necessary to borrow information across subjects to recover the temporal trend of the network edge parameters for individual subjects. Doing so requires us to restrict the temporal trajectory within a time window (of length 500 days) covering all subjects. The proportion of subjects with scans outside of the 500 day window is not sufficient to recover the temporal trajectory outside of the window, and thus we restrict our analysis to the 319 rs-fMRI scans falling within it. The advantage of doing this is that we are able to use the function-on-scalar regression model where the response is the temporal profile of the network edge parameters. 

The function-on-scalar regression model considers the association of the temporal trend of the network edge parameters with genetic variables, while the linear mixed effects model only focuses on the association of the network edge parameters with the genetic variables. The mixed-effects model simply treats the network edge parameters at multiple time points as repeated measurements and accounts for clustering, which ignores the time order and trend of the network edge parameters. We view this as a key aspect of the longitudinal data. Therefore, the function-on-scalar model may be able to find a stronger association between the network edge parameters with genetics as it accounts explicitly for time ordering (and not just clustering) in the network response. Thus after preprocessing the data in this way we are left with 319 scans obtained from 111 subjects for the longitudinal analysis. 

MRI for these subjects are collected at 3T. MRI data are downloaded with permission from the ADNI. All images were acquired on 3.0 Tesla Philips MRI scanners across 10 North American acquisitions sites according to the standardized ADNI protocol. Whole-brain anatomical MRI scans were acquired sagittally, with a T1-weighted magnetization-prepared rapid acquisition with gradient echo (MPRAGE) sequence, with the following parameters: 1.2 mm slice thickness, 256 by 256 by 170 acquisition matrix, echo time (TE) of 3 ms, in-plane voxel dimension of 1 mm$^{2}$, repetition time (TR) of 7 ms, and flip angle of 9 degrees. Functional MRI scans were obtained during resting-state; participants were instructed to lay quietly in the scanner with their eyes open. Resting state fMRI scans were obtained with a T2*-weighted echo-planar imaging sequence with the following parameters: 140 volumes, 64 by 64 by 48 acquisition matrix (voxel size = 3.3 mm$^{3}$), TE of 30 ms, TR of 3000 ms, and flip angle of 80 degrees.

The freely available software package PLINK (Purcell et.al., 2007) is used for genomic quality control and preprocessing. The genetic data are SNPs from non-sex chromosomes, i.e., chromosome 1 to chromosome 22. SNPs with minor allele frequency less than 5\% are removed as are SNPs with a Hardy-Weinberg equilibrium p-value lower than $10^{-6}$ or a missing rate greater than 5\%. After preprocessing we are left with 1,220,955 SNPs for each of 111 subjects for relating genetic data to effective connectivity and 1,220,955 SNPs for each of an additional 663 subjects for selecting a disease-constrained subset of SNPs. 

\subsection{rs-fMRI Data Preprocessing and Network Estimation}

The fMRI and anatomical data are pre-processed using a combination of the FSL software (available at \url{http://fsl.fmrib.ox.ac.uk/fsl/fslwiki/}) and the SPM12 software (available at \url{http://www.l.ion.ucl.ac.uk/spm/software/spm12/}).  Non-brain tissue in the raw T1 images is removed using the automated Brain Extraction Tool (Smith, 2002), followed by manual verification and optimization for each subject. Blood-Oxygen-Level Dependent (BOLD) image data preprocessing is performed in FSL's FEAT as follows: each functional image is motion corrected and registered to their high-resolution T1 structural image that is linearly registered to standard stereotaxic space using a 12 degree-of-freedom transformation. A non-linear registration of the structural image to standard stereotactic space is also applied to account for potential local deformations in brains of the patient group. Each subject's imaging data are normalized to a standardized space defined by an MNI template brain. We conducted manual checking for correct normalization of every image. 

Given a set of $R$ brain regions of interest, DCM in the case of fMRI models the haemodynamic response over these regions through a nonlinear state-space formulation with a model allowing for interaction between regions and with model parameters that characterize effective connectivity and, when relevant, how this connectivity is modulated by experimental inputs. In the case of resting-state fMRI with no experimental inputs, the model can be expressed as (see, Razi et al., 2017)
\begin{equation}
\label{DCM_state}
\dot{\vx}(t) = \vA \vx(t) + \vv(t)
\end{equation}
$$
\vy(t) = h(\vx(t), \vtheta) + \ve(t),
$$
where $\vx(t) = (x_{1}(t),\dots,x_{R}(t))'$ are latent variables used to represent the states of neuronal populations at some time $t$ and $\dot{\vx}(t)$ is a time-derivative defining a differential equation approximating neuronal dynamics, with the $R \times R$ matrix $\vA$ approximating effective connectivity to first-order; $h(\vx(t), \vtheta)$ is a nonlinear mapping from hidden neuronal states to the observed haemodynamic response also depending on parameters $\vtheta$ (see, e.g., Friston, 2007, for details on the form of this nonlinear mapping); $\vy(t) = (y_{1}(t),\dots,y_{R}(t))'$ with $y_{j}(t)$ being a summary of the response obtained from all voxels within region $j$; $\vv(t)$ and $\ve(t)$ represent neuronal state noise and measurement noise respectively. 

For resting-state fMRI, the DCM can be fit in the time-domain using Bayesian filtering based on a mean-field variational Bayes approximation (see, e.g., Li et al., 2011) which involves inference on both model parameters and latent states. Alternatively, the model can be fit in the spectral domain using an approach known as spectral DCM (Friston et al., 2014). The latter approach involves relating the theoretical cross spectra associated with the dynamic model to the sample cross spectra in order to estimate parameters. Thus, it is somewhat akin to a method of moments approach. More specifically, Friston et al. (2014) assume a parameterized power law form for the spectral densities of the noise terms in the state-space model and then express the empirical cross spectra as the sum of the theoretical cross spectra and measurement error. This formulation then yields a likelihood for the observed cross spectra statistic depending on the time-invariant parameters but not depending on the latent variables $\vx(t)$. This likelihood for the summary statistic is then combined with a prior distribution for the model parameters and an approximation to the associated posterior distribution for these parameters is obtained using variational Bayes. Razi et al. (2015) report simulation results that demonstrate estimators obtained from spectral DCM having higher accuracy (in the sense of mean-squared error) than those obtained from stochastic DCM. In addition, the former has a higher computational efficiency since estimation of the latent states is not required. We use this approach to estimate effective connectivity networks within the DMN.

The DMN includes the posterior cingulate cortex/precuneus (PCC), medial prefrontal cortex (MPFC), bilateral inferior parietal lobule (IPL), and other regions including the inferior temporal gyrus. To estimate effective connectivity within the regions of the DMN depicted in Figure 1 and Table 1, we use spectral DCM as implemented in SPM12. BOLD time series from the DMN regions of interest are obtained by extracting time series from all voxels within an 8mm radius of the associated MNI coordinate, and then applying a principle component analysis and extracting the first eigenvariate. For simplicity, the information contained in the other eigenvariates is not considered in our analysis so that the result is a single representative time series for each region of interest. This procedure is repeated to obtain a collection of four (DMN4) or six (DMN6) time series for each subject. An example of the resulting time-series data for a single subject over DMN6 is depicted in Figure 2. 

A 16-parameter (DMN4) or 36-parameter (DMN6) graph with weights representing effective connectivity between and within regions is then estimated for each subject. This graph is based on an estimate of the parameter $\vA$, an $R \times R$ non-symmetric matrix, in equation (\ref{DCM_state}) obtained from spectral DCM. We fit the DCM in SPM12 with the option of one state per region and with the model fit to the cross spectral density (which corresponds to spectral DCM). The imaging preprocessing pipeline is summarized in Figure 3. 

\section{Selection of the Disease-Constrained Set of SNPs}\label{Study2}

Beginning with the 1,220,955 SNPs discussed in Section 2, we conduct a genome-wide association study (GWAS) with the goal of identifying a smaller subset of SNPs that are potentially associated with disease (CN/MCI/AD). This subset serves as a constraint in subsequent analysis examining effective brain connectivity and it is selected out-of-sample.  That is to say, the selection of the constrained SNP subset does not involve the 111 subjects with rs-fMRI data but rather is obtained from taking the original sample of 774 subjects having genome-wide data and removing those 111 subjects with rs-fMRI. The remaining 663 ADNI2 subjects have genome-wide data only and we conduct the GWAS on the data from these subjects in order to identify a smaller subset of SNPs that are potentially associated with disease (CN/MCI/AD). This yields a subset of top 100 SNPs which we then relate to the longitudinal rs-fMRI effective connectivity networks using the 111 subjects with rs-fMRI data. This approach has two primary advantages in producing a more reliable subset of SNPs than within sample selection:
\begin{enumerate}
	\item The disease-constrained set is based on a larger sample size of 663 subjects compared with 111 subjects.
	\item Selecting the subset out-of-sample avoids double usage of the data in selecting the priority subset of SNPs and in relating that subset to longitudinal rs-fMRI.
\end{enumerate}

A multinomial logistic regression with disease category as the response is fit for each SNP to assess that SNP's marginal association with disease after adjusting for covariates representing age, sex, handedness, and education. SNPs are included in the model as the number of a particular allele so that a SNP's effect on the log-odds ratio is additive. We sort the SNPs by the resulting p-values from a likelihood ratio test, where the null hypothesis corresponds to the case where the probability distribution of disease does not depend on the given SNP. A subset of the top 100 SNPs is selected based on this ranking. The distribution of p-values by chromosome and the cut-off for selecting the best subset of 100 SNPs is depicted in Figure \ref{fig:figure1}.  Each of the 100 SNPs in the selected constrained subset has a p-value below $7.5 \times 10^{-5}$, which represents the cut-off indicated in Figure \ref{fig:figure1}. 

While Table 1 indicates an APOE signal in our data from the $\epsilon$4 allele of the gene, we note that there are no APOE related SNPs in our constrained subset of the top 100. The highest ranking APOE related SNP is rs4802234 which has a p-value of $6.9 \times 10^{-4}$. Thus our choice to use a p-value threshold of $7.5 \times 10^{-5}$ in subsequent analyses eliminates the highest ranking APOE related SNP, even though there appears to be some evidence of an APOE signal in the data. We note that we have defined APOE related SNPs as those within a 1 million base pair range of APOE and their p-values ranging between $6.9 \times 10^{-4}$ to 0.999. Despite this, all subsequent regression models that we use will include a variable representing APOE genotype (coded as number of $\epsilon$4 alleles) in order to account for its known importance and to adjust for its potential association with effectivity brain connectivity.

%Having identified the constrained subset, we jointly assess the effect of these SNPs using a symmetric multinomial logistic regression with LASSO penalty and with disease (CN/MCI/AD) as the response. We fit a model including all of the SNPs in the constrained-subset only and another model including the same SNPs along with covariates representing age, sex, handedness, and education and the disease status (CN/MCI/AD) as the response. The glmnet software (Friedman et al., 2010) is used to fit the LASSO penalized symmetric multinomial logistic regression model, where each class is represented by a linear model on the log-scale and the penalty allows for redundancies so that all levels of the response have an associated parameter vector in the model.  The estimated regularization path for both models is shown in Figures 5 and 6. 
%
%Examination of Figure 5 shows two SNPs on chromosome 11, kgp239829 and rs4910743 both in high linkage (correlation 0.94) that stand out from the other SNPs in terms of their coefficient 2-norms when related to each category. Examination of Figure 6 shows that these same SNPs continue to have regularization paths that stand out among other genetic variables when  age, sex, handedness, and education are included in the model as well. We note that the p-value for kgp239829 obtained from the marginal analysis relating this SNP to disease is p-value $= 4.876 \times 10^{-5}$ and it obtains a rank of 82 out of 1,220,955. {\bf Refer to table showing SNPs correlated with kgp239829.}

\section{Resting-State Effective Brain Connectivity by Genetics}\label{Study3}

\subsection{Longitudinal Analysis with Four DMN Regions}

We conduct the longitudinal analysis using both a linear mixed effects model incorporating subject-specific random effects as well as a function-on-scalar regression. In each case the response is a network connection between two regions of interest estimated from an rs-fMRI scan using DCM, and we observe a longitudinal sequence of such connections for each subject with the number of repeated measurements ranging from 1 to 4 with a median of 3 and a third-quartile of 4.

For a given connection, we will fit a regression model relating that connection to a single SNP, for each of the 100 SNPs in the disease-constrained set. These regression models for the time-sequence of connection values also include covariates representing age, sex, right/left hand, education and APOE$\epsilon$4. With 16 possible connections a total of $16 \times 100 =1600$ models are fit and in each case a test comparing the model with and without the SNP from the constrained-set is conducted. The resulting set of $1600$ p-values arising from this mass-univariate longitudinal approach are adjusted for multiplicity using an FDR adjustment (Benjamini and Hochberg, 1995) in order to produce for each edge-SNP pair a q-value. If a q-value $=\alpha$ cut-off is selected then among all rejections based on this cut-off, there are about $100 \times \alpha$  \% that are falsely rejected. In this exploratory analysis we focus primarily on the rank of each edge-SNP pair by the p-value but we present the q-value as well.

The linear mixed effect (LME; Kuznetsova et al., 2017) model has the form
\begin{align*}
Y_{ij} &=  \mu + X_{i}\beta + \vZ_{i}^{T} \valpha + b_{i} + \epsilon_{ij}
\end{align*}
where $Y_{ij}$ is the network edge weight estimated from the $j^{th}$ rs-fMRI scan of the $i^{th}$ subject, $i = 1, 2, ...,111$, $j = 1, 2, ...,m_i$; $X_{i}$ is the SNP coded additively in the model; $\vZ_{i}$
contains the remaining covariates with coefficient vector $\valpha$, $b_{i}$ is a subject-specific random effect and  $\epsilon_{ij}$ is an error term with the random effect and errors assumed Gaussian and independent. The likelihood ratio test is conducted for the hypothesis test  $H_0: \beta =0$ against $H_1: \beta \neq 0$ and the corresponding p-value obtained.  

In addition to obtaining a p-value from the standard $\chi^{2}_{1}$ asymptotic null distribution for the likelihood ratio statistic we also apply a parametric bootstrap. For the linear mixed effects model the parametric bootstrap procedure proceeds by first fitting both the null (excluding the SNP) and alternative (including the SNP) model using maximum likelihood and the observed value of the likelihood ratio statistic is obtained. Data are then generated under the null model and both the null and alternative model are fit to each simulation replicate and a realization of the test statistic under the null hypothesis is obtained. This is repeated for $n_{sim} = 300,000$ replicates and the bootstrap p-value is obtained as the proportion of null test statistic realizations as large or larger than the observed test statistic.

We apply the PACE approach for sparse functional data to obtain the estimated trajectories of network edge weights over time $t$, $\widehat{Y}_{i}(t)$. By applying PACE, the estimated trajectory of each network edge weight can be obtained and expressed as
\begin{align*}
\widehat{Y}_{i}(t) &= \mu(t) + \sum_{j=1}^{J} \xi_{ij}\phi_{j}(t), \\ 
\widehat{\pmb{Y}}(t) &= \mu(t) +  \pmb{\xi}_{n\times J} \pmb{\Phi}(t)_{J \times 1},
\end{align*}
where $\mu(t)$ represents the mean function common to all $n$ subjects, $$\xi_{ij} = \int\{Y_{i}(t) - \mu(t)\}\phi_{j}(t) dt$$ is the $j$-th functional principal component (FPC) score for $Y_{i}(t)$, with $\phi_{j}(t)$ being the corresponding eigenfunction, and $J=3$ since three functional principal components explain over 95\% of the total of variation. 

The function-on-scalar regression model has the form
    \begin{equation*} \label{eqn:FM}
     Y_{i}(t) = X_{i}\alpha(t) + \vZ_{i}^{T}\vbeta(t) + \epsilon_{i}(t),
     \end{equation*}
where $X_{i}$ is the SNP with corresponding regression function $\alpha(t)$, $\vZ_{i}$ is a vector holding covariates representing age, sex, right/left hand, education and APOE$\epsilon$4 with corresponding regression function $\vbeta(t)$ and the error term $\epsilon_{i}(t)$ is a random error processes. As the raw data $Y_{i}(t)$ is sparse we replace it in the regression model with the PACE-estimated trajectory of each network edge weight. We test $H_{0}:\, \alpha(t) = 0$ corresponding to the model that excludes the genetic marker using an  \textit{F}-test. As with our use of linear mixed models we apply a mass univariate approach and apply the model to all $16 \times 100 =1600$ possible combinations of network edge and SNP, and apply an FDR correction for multiplicity. We note that the degrees of freedom for the F-distribution needs to be carefully computed. In the F-test for FSR the degrees of freedom are setup by applying the idea of Satterthwaite's approximation described in Shen and Faraway (2004). 

In addition to obtaining p-values from the F-distribution we also apply a parametric bootstrap. To the best of our knowledge, there is no existing package or function in R that can be used for simulation with the function-on-scalar regression (FSR) model. We therefore propose the following procedure to generate functional data for bootstrapping the test statistics under the null and its corresponding p-value for hypothesis testing the SNP regression function in FSR models.

The two FSR models considered in the hypothesis test for the SNP effect are:
\begin{enumerate}
	\item Null model: $Y_{i}(t) = \sum_{j=1}^{J}Z_{ij}\beta_{j}(t) + \epsilon_{i}(t)$, where the SNP regression function $\alpha(t) = 0$
	\item Alternative hypothesis: $Y_{i}(t) =  X_{i}\alpha(t) + \sum_{j=1}^{J}Z_{ij}\beta_{j}(t) + \epsilon_{i}(t)$, where $\alpha(t) \ne 0$ 
\end{enumerate}
For each bootstrap sample we simulate the residual function $\epsilon_{i}(t)$ with the empirical mean and covariance functions of the estimated residual function $\hat{\epsilon}_{i}(t)$ of the FSR model under the null. We note that in the FSR model, each residual term $\epsilon_{i}(t)$ is assumed to arise from a Gaussian process with mean zero and covariance function $r(s,t)$. We then generate the functional response $Y_{i}(t)$ using the null FSR model with the original scalar covariates $Z_{ij}$, their corresponding estimated functional coefficients $\hat{\beta}_{j}(t)$, and the simulated residual functions from the previous step. We then re-fit the two FSR models from the null and alternative hypotheses to the functional responses simulated under the null  and compute the F-statistic as a realization under the null. This procedure is repeated to obtain $n_{sim} = 10,000$ replicates and the bootstrap p-value is obtained as the proportion of null test statistic realizations as large or larger than the observed test statistic.

The LME assumes that the effective-connectivity is time-invariant and all of the session-specific fluctuations are captured by the noise term in the LME. The FSR model on the other hand allows for session-specific fluctuations through the time-varying regression functions. This allows the FSR analysis to be potentially more powerful since all of the fluctuations are not allocated to the noise term. An important point is that LME and FSR are approaches that make fundamentally different assumptions about the data generating mechanism.

In Tables 3 and 4 we display the top ten ranked edge-SNP associations with the smallest p-values for DMN4 as obtained from both LME and FSR. The bootstrap p-values (B-p-values) as well as the FDR adjusted q-values are also indicated in the tables. As an exploratory analysis of these results we intersect the top 20 edge-SNP associations out of 1600 (based on the p-value) from LME with the top 20 edge-SNP associations out of 1600 (based on the p-value) from FSR. Our rationale for doing this is that both LME and FSR are based on very different modelling assumptions with different model fitting procedures used in each. Therefore, signals that appear as highly ranked from both approaches simultaneously may represent genuine signal in the data since it is unlikely that noise will be stable to the different assumptions and procedures applied to the data. This sort of stability to differing modelling assumptions provides an alternative potentially useful approach to detecting signal in small-sample high-dimensional settings such as those commonly encountered in imaging genetics. Intersecting the top 20 associations from both methods produces a very clear pattern of 8 associations that are presented in Table 5. The pattern reveals a potential signal involving connections LIPC $\rightarrow$ RIPC, LIPC $\rightarrow$ PCC and PCC $\rightarrow$ PCC and SNPs from chromosome 9 and 13.

%The results arising from applying the linear mixed model approach and the function-on-scalar regression approach are presented in Tables 3 and 4, respectively, which show results corresponding to FDR adjusted p-values less than 0.1. Both LME and FSR identify the same potential association between the connection from the posterior cingulate cortex to the left intraparietal cortex and a genetic signal represented by SNP rs2602615/kgp3833472 on chromosome 2 (FDR adjusted p-value = 0.092 for FSR; FDR adjusted p-value = 0.096 for LME). Both LME and FSR identify the same potential association suggesting a strong degree of robustness to model assumptions. 

\subsection{Longitudinal Analysis with Six DMN Regions}
We apply the same approaches to the networks estimated using spectral DCM for the network of 6 regions listed in Table 1. In this case the models are applied to $36 \times 100 =3600$ possible combinations of network edge and SNP, and a p-value, B-p-value and and FDR corrected q-value is obtained to adjust the p-value for multiplicity. 

In Tables 6 and 7 we rank the ten associations with the smallest p-values for DMN6 as obtained from LME and FSR. We again conduct an exploratory analysis of these results by intersecting the top 20 edge-SNP associations out of 3600 (based on the p-value) from LME with the top 20 edge-SNP associations out of 3600 (based on the p-value) from FSR. For DMN6 we see seven intersecting top 20 associations involving self connections (regions modulating their own activity) for MPFC, LIT and RIT involving SNPs on chromosomes 11 and 6. Regions in chromosome 6 have been previously implicated for Late-Onset Alzheimer Disease (Naj et al., 2010). There are also two intersecting top 20 associations between PCC $\rightarrow$ RIPC and SNPs on chromosome 18, and an intersecting association between LIT $\rightarrow$ LIPC and SNP rs6021246 on chromosome 20. The stability of these associations makes them of potential interest as signal between the larger network and disease-constrained SNPs though we emphasize that these results are exploratory. While we see a non-trivial change in the results when moving from DMN4 to DMN6 this can be explained simply by noting that many of the higher ranked SNP-edge pairs in DMN6 involve the two added nodes corresponding to ROIs RIT and LIT.

Tables 3, 4, 6 and 7 make comparisons between the p-values obtained from the $\chi^{2}_{1}$ and $F$-distribution and the B-p-values obtained from the parametric bootstrap procedures developed for both LME and FSR. Generally, the p-values are inline with the B-p-values and the comparison from the four tables suggest that the asymptotic distributions provide an adequate approximation for the case of these data. To investigate this further we plot in Figures 5 and 6 the histograms of the 3600 (1600 in the case of DMN4) p-values obtained from all of the tests relating SNPs to effective connectivity as obtained from the $\chi^{2}_{1}$ and $F$-distribution. As most SNPs should be under the null, adequate performance of the asymptotic distributions should be indicated by uniform distributions. In all four cases (DMN4-LME, DMN6-LME, DMN4-FSR, DMN6-FSR) the distributions of the p-values look reasonably uniform which, in addition to the comparisons between p-values and B-p-values made in Tables 3, 4, 5 and 6, further suggests that the asymptotic null distributions are preforming adequately in the cases considered here. 

%For DMN6, LME does not identify any edge-SNP combinations with FDR adjusted p-value $\le$ 0.1. In contrast, FSR may be more powerful as it incorporates the information on time-ordering (on the time scale of fMRI scans) and the corresponding results are presented in Table 5. Examining Table 5, the difference in the results we see in moving from DMN4 to DMN6 can be explained by the heavy involvement of the right hippocampus (RH) and right inferior temporal (RIT) region which are nodes in DMN6 but not in DMN4. These regions are involved in seven of the ten results reported in Table 5.
%
%FSR suggests potential associations between the connections from the left intraparietal cortex to the right inferior temporal cortex and SNP kgp7160266 from chromosome 12 (FDR adjusted p-value = 0.038); the connection from the left intraparietal cortex to the right intraparietal cortex and SNP rs7190071 from chromosome 16 (FDR adjusted p-value = 0.040), the connection from the right hippocampus to the posterior cingulate cortex and SNP kgp10652659 from chromosome 11 (FDR adjusted p-value = 0.050) and the connection from the right hippocampus to right intraparietal cortex and SNP rs7617199 from chromosome 3 (FDR adjusted p-value = 0.050).

\section{Discussion}

We have examined the association between longitudinal effective brain connectivity and genetics based on an empirically derived disease-constrained set of SNPs using LME and FSR, with networks estimated from rs-fMRI time series using spectral DCM. Our analysis is based on obtaining a disease-constrained subset of SNPs based on the outcome of the out-of-sample GWAS in Section 3, the purpose being to focus attention on potential associations that are empirically most relevant for disease. We have examined both a 4-region and a 6-region network depicted in Figure 1 and Table 1. 

We have examined both a 4-region and a 6-region network so as to obtain a more complete picture of the data. Indeed, our results from DMN4 obtained after intersecting the top associations reveal an interesting pattern that would not have been noticed if only the larger DMN6 network was analyzed where many of the top-ranked edge-SNP pairs involve LIT and RIT. As part of future work we intend to increase the number of regions to a much larger size (e.g. networks where the number of nodes is an order of magnitude greater and spanning multiple known resting-state networks) and examine measures of network topology as opposed to network edge values as a potential useful low-dimensional phenotype. 

After adjusting for multiple comparisons none of the associations examined have a q-value below 0.1. By intersecting the top 20 associations and focussing on stability to the assumptions made by LME and FSR we do observe some patterns that may be of interest from an exploratory point of view. The most interesting patterns can be summarized as follows:
\begin{enumerate}
	\item DMN4: LIPC $\to$ RIPC and three SNPs on chromosome 9 (rs13287994, kgp4931190, kgp9433690) exhibit top 20 associations that are stable under LME and FSR
	\item DMN4: LIPC $\to$ PCC and two SNPs on chromosome 9 (kgp9433690, rs13287994) exhibit top 20 associations that are stable under LME and FSR. The large genomic screen conducted by Pericak-Vance et al. (2000) identified potential AD-related regions on chromosome 9.
	\item DMN4: PCC $\to$ PCC and three SNPs on chromosome 13 (rs9317920, kgp12216228, rs1935110) exhibit top 20 associations that are stable under LME and FSR
	\item DMN6: PCC $\to$ RIPC and two SNPs on chromosome 18 ( rs949200, kgp8588069) exhibit top 20 associations that are stable under LME and FSR
	\item DMN6: Several self-connections involving the MPFC, LIT, RIT on chromosomes 6 and 11. Regions in chromosome 6 have been previously implicated for Late-Onset Alzheimer Disease (Naj et al., 2010).
\end{enumerate}

Our comparisons of the parametric bootstrap procedures for LME and FSR to the p-values obtained from the $\chi^{2}_{1}$ and $F$-distribution suggest that the asymptotic distributions are performing adequately for the data considered here. This is evident both in the comparisons made in Tables 3, 4, 6 and 7 as well as in Figures 5 and 6. Further investigation of this issue, in particular for testing regression functions in FSR, is an avenue for future work.

Our current analysis suggests follow-up analyses looking at high resolution anatomical and diffusion tensor imaging and structural connectivity in a separate sample of subjects with such an analysis focussed on a small number of targeted SNPs suggested by the current analysis. From a methodological perspective, our approach of determining the disease-constrained set first and subsequently examining relationships with effective brain connectivity motivate the development of a joint model examining both disease and longitudinal effective connectivity simultaneously with a model structure that allows for joint variable (SNP) selection. 

While connectivity may be related to the degeneration of gray matter our regression analyses have not incorporated gray matter density as a covariate in either the LME or FSR models. In future work we will look to reporting on an analysis that includes grey matter density in the ROIs for each subject as well as examine a much larger set of ROIs including bilateral hippocampi and the combination of structural and effective connectivity. 

We note again that our analysis summarizes ROI specific activity by extracting the time series at all voxels within a sphere having radius 8mm around an associated MNI coordinate for the ROI. A principal component analysis is applied to these time series and the first eignenvariate is used as a region-specific summary. Subsequent analysis may consider additional eigenvariates which may hold useful information representing effective connectivity. 

While our analysis has focussed on hypothesis tests examining individual SNPs, an alternative is to cluster the SNPs using hierarchical clustering and fit the LME and FSR models to the first genetic principal component scores of each cluster. In supplementary analyses not reported in Section 4 we have clustered the top 100 SNPs into 22 clusters based on hierarchical clustering to facilitate dimension reduction prior to relating genetics to effective brain connectivity. After applying SNP clustering and relating the first principal component score of the SNP clusters to effective connectivity network edges both LME and FSR find simultaneously a rank 1 association between a specific cluster of 12 SNPs (rs2465362, kgp9051645, kgp5238984, kgp9055011, kgp3071374, rs9388153, rs2860410, kgp22784175, rs7335304, rs9317920, kgp12216228, rs1935110) and the self connection PCC $\rightarrow$ PCC (p-value = $3.52 \times 10^{-4}$, q-value = 0.0502 for LME; p-value = $3.05 \times 10^{-4}$, q-value = 0.0585 FSR). This potential association between the resting-state activity of the posterior cingulate cortex and genetics is further corroborated by our results reported in Table 5 where individual associations between three SNPs in this cluster and PCC $\rightarrow$ PCC are also reported in the high ranking intersection set obtained from combining LME and FSR.

The neuropathological mechanisms that underly the effective connectivity observed in this study could be related to the usual hallmarks of AD and MCI. For example, amyloid plaques, neurofibrillary tangles and/or structural neurodegeneration may underly effective connectivity as well. Therefore, our current results suggest follow-up studies incorporating measures of amyloid beta and tau (as measured in CSF or by PET imaging) and measures of brain structure (as measured by high resolution anatomical and diffusion tensor imaging) and examination of the relationship between these measures and the genetic variables suggested by this analysis.

%%%%%%%%%%%%%%%%%%%%%%%%%%%%%%%%%%%%%%%%%%%%%%%%%%%%%%%%%%%%%%%%%%%%%%%%%%%%%%%%%%%%%
%
%     please remove the " % " symbol from \centerline{\includegraphics{fig01.eps}}
%     as it may ignore the figures.
%
%%%%%%%%%%%%%%%%%%%%%%%%%%%%%%%%%%%%%%%%%%%%%%%%%%%%%%%%%%%%%%%%%%%%%%%%%%%%%%%%%%%%%%

\section*{Acknowledgements}
Research is supported by funding from the Natural Sciences and Engineering Research Council of Canada (NSERC) and the Canadian Statistical Sciences Institute. F.S. Nathoo holds a Tier II Canada Research Chair in Biostatistics for Spatial and High-Dimensional Data. This research is also partially supported by an internal research grant from the University of Victoria. Research was enabled in part by support provided by WestGrid (www.westgrid.ca) and Compute Canada (www.computecanada.ca). The authors acknowledge support with data storage and computing issues from WestGrid and Compute Canada. Data collection and sharing for this project was funded by the Alzheimer's Disease Neuroimaging Initiative (ADNI) (National Institutes of Health Grant U01 AG024904) and DOD ADNI (Department of Defense award number W81XWH-12-2-0012). ADNI is funded by the National Institute on Aging, the National Institute of Biomedical Imaging and Bioengineering, and through generous contributions from the following: AbbVie, Alzheimer's Association; Alzheimer's Drug Discovery Foundation; Araclon Biotech; BioClinica, Inc.; Biogen; Bristol-Myers Squibb Company; CereSpir, Inc.; Cogstate; Eisai Inc.; Elan Pharmaceuticals, Inc.; Eli Lilly and Company; EuroImmun; F. Hoffmann-La Roche Ltd and its affiliated company Genentech, Inc.; Fujirebio; GE Healthcare; IXICO Ltd.; Janssen Alzheimer Immunotherapy Research \& Development, LLC.; Johnson \& Johnson Pharmaceutical Research \& Development LLC.; Lumosity; Lundbeck; Merck \& Co., Inc.; Meso Scale Diagnostics, LLC.; NeuroRx Research; Neurotrack Technologies; Novartis Pharmaceuticals Corporation; Pfizer Inc.; Piramal Imaging; Servier; Takeda Pharmaceutical Company; and Transition Therapeutics. The Canadian Institutes of Health Research is providing funds to support ADNI clinical sites in Canada. Private sector contributions are facilitated by the Foundation for the National Institutes of Health (www.fnih.org). The grantee organization is the Northern California Institute for Research and Education, and the study is coordinated by the Alzheimer's Therapeutic Research Institute at the University of Southern California. ADNI data are disseminated by the Laboratory for NeuroImaging at the University of Southern California. 

\pagebreak

\pagebreak

\pagebreak

	\begin{figure}[h]
		\centering
		\includegraphics[width=0.75\textwidth]{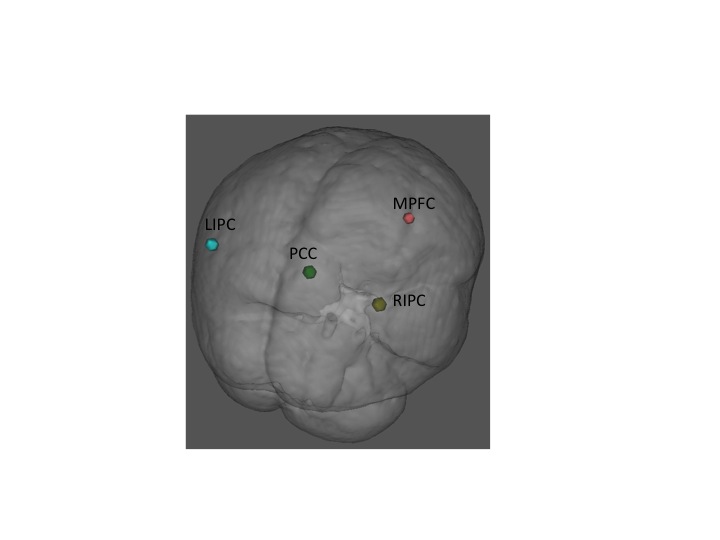}
		\vspace{-1cm} 
		\caption{The locations of the four regions within the default mode network (DMN) examined in our DMN4 study: the medial prefrontal cortex (MPFC), the posterior cingulate cortex (PCC), the left and right intraparietal cortex (LIPC and RIPC) with MNI coordinates MPFC (3, 54, -2), the PCC (0, -52, 26), LIPC (-50, -63, 32) and RIPC (48, -69, 35).}
		\label{fig:figure2}
	\end{figure}

\pagebreak

\section*{Figures}

		\begin{figure}[h]
		\centering
		\includegraphics[width=.85\textwidth]{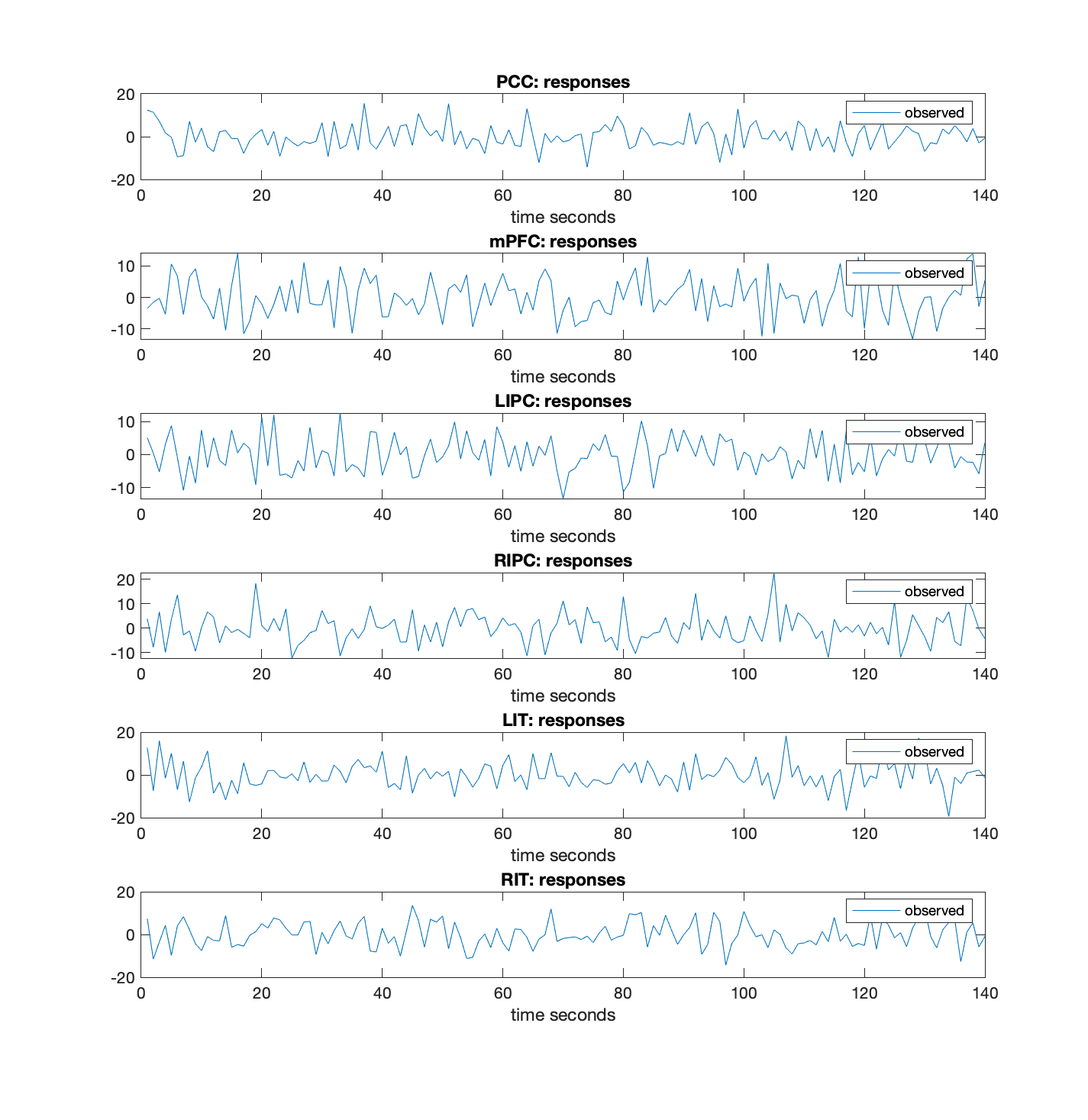}\\
		\caption{An example of the rs-fMRI data used to estimate the effective connectivity network for a single subject from the six regions of interest (PCC, MPFC, LIPC, RIPC, LIT, RIT).}	
		\end{figure}	

\pagebreak

	\begin{figure}[h]
		\centering
		\includegraphics[width=0.35\textwidth]{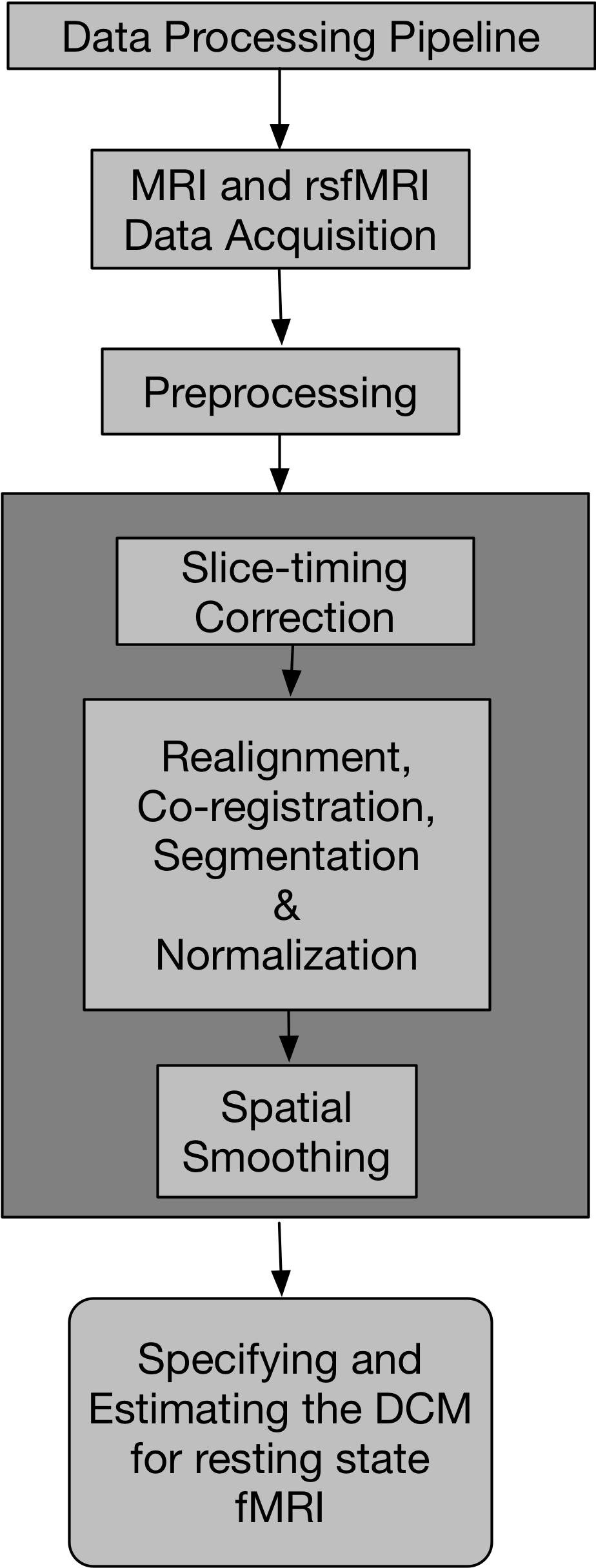}
		%\vspace{-1cm} 
		\caption{The neuroimaging data preprocessing pipeline used in our study.	Slice-timing correction as implemented in SPM12 is applied to all of the functional images. All slices of one volume are interpolated in time to the reference slice (reference slice = 24 out of 48). Spatial smoothing is performed in SPM12 where we convolve image volumes with a spatially stationary Gaussian filter (i.e. a Gaussian kernel ) of $8\times 8 \times 8$ mm$^3$ full width half max (FWHM).}
		\label{fig:figure2}
	\end{figure}

\pagebreak

	\begin{figure}[h]
		\centering
		\includegraphics[width=0.75\textwidth]{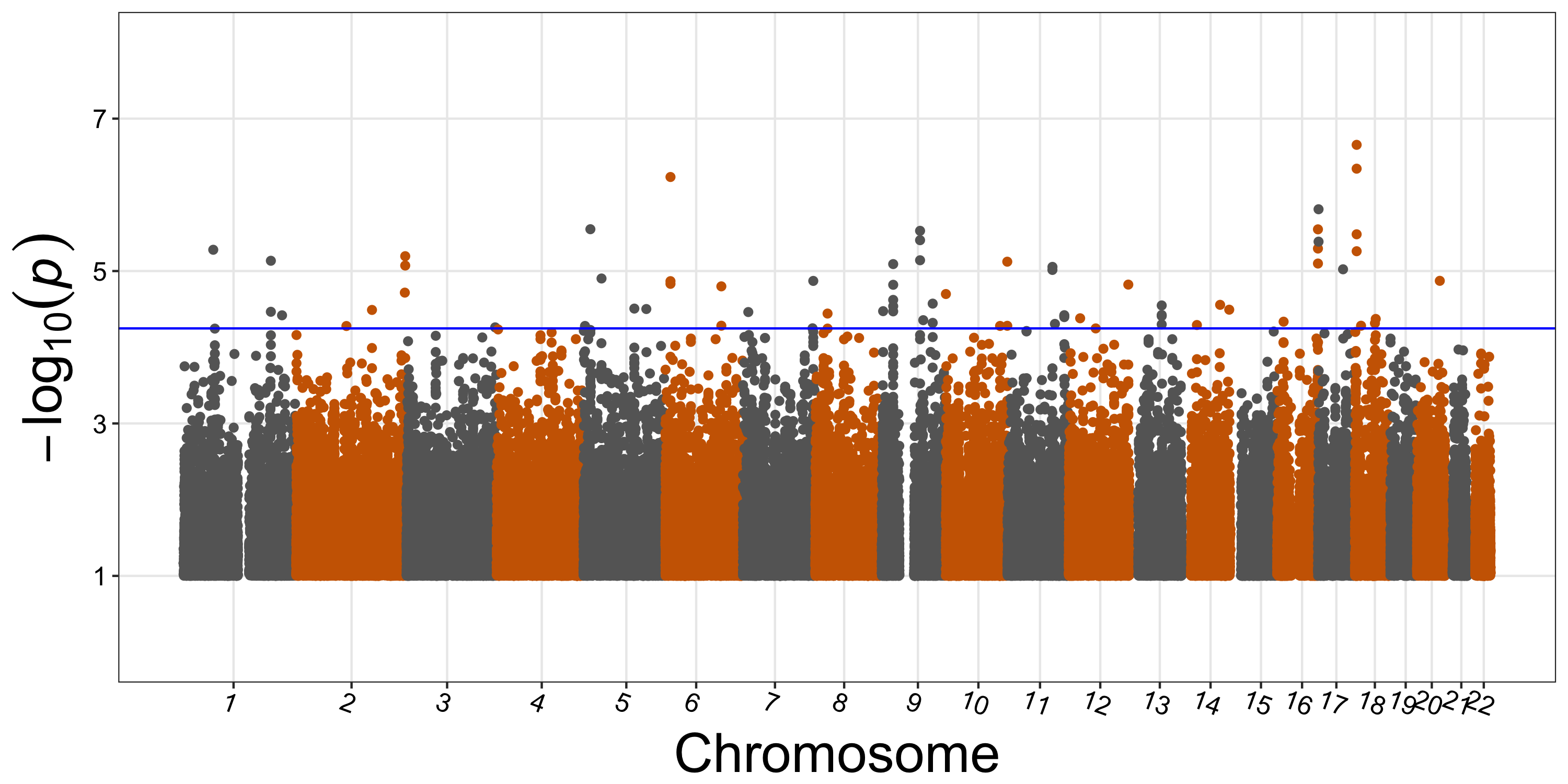}
		\caption{The p-values associating disease status with SNPs adjusting for age, sex, handedness, and education. The blue line represents the cutoff used to obtain the top 100 SNPs which corresponds to a p-value threshold of $7.5 \times 10^{-5}$.}
		\label{fig:figure1}
	\end{figure}

\pagebreak

	\begin{figure}[h]
		\centering
	\includegraphics[scale=0.45]{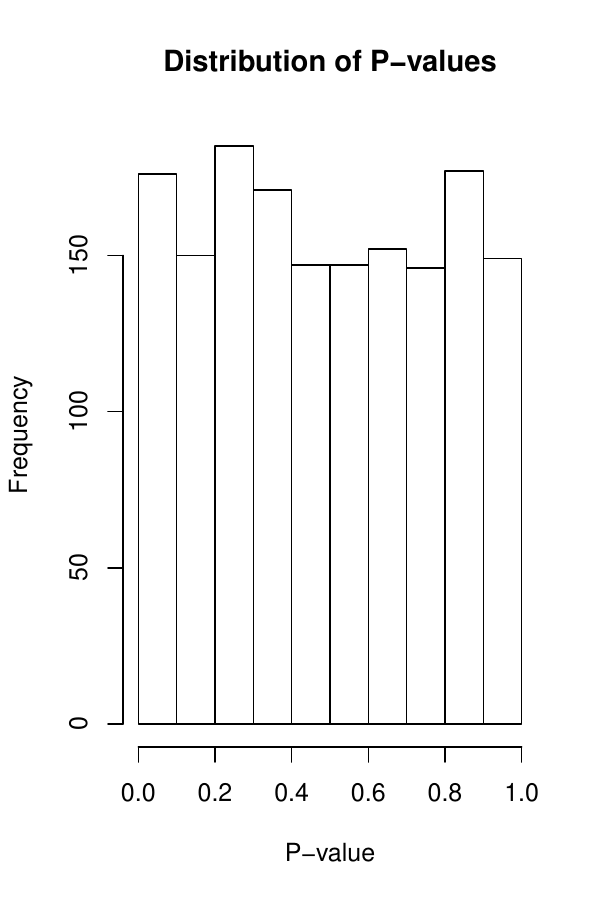}
	\includegraphics[scale=0.45]{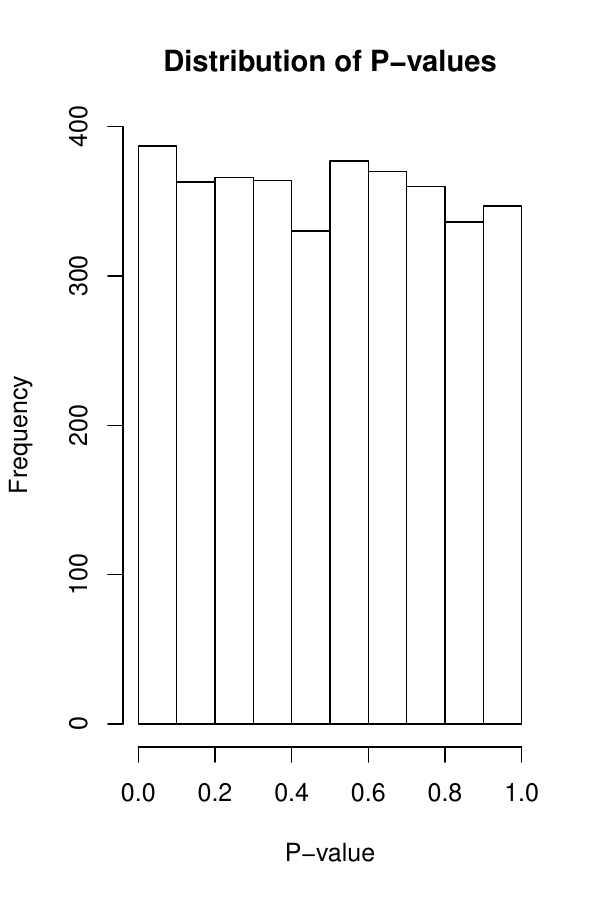} 
	\caption{Distribution of p-values obtained from the $\chi^{2}_{1}$ null distribution for LME for DMN4 (left) and DMN6 (right).}
		\label{fig:figure1LME}
	\end{figure}

\pagebreak

	\begin{figure}[h]
		\centering
	\includegraphics[scale=0.3]{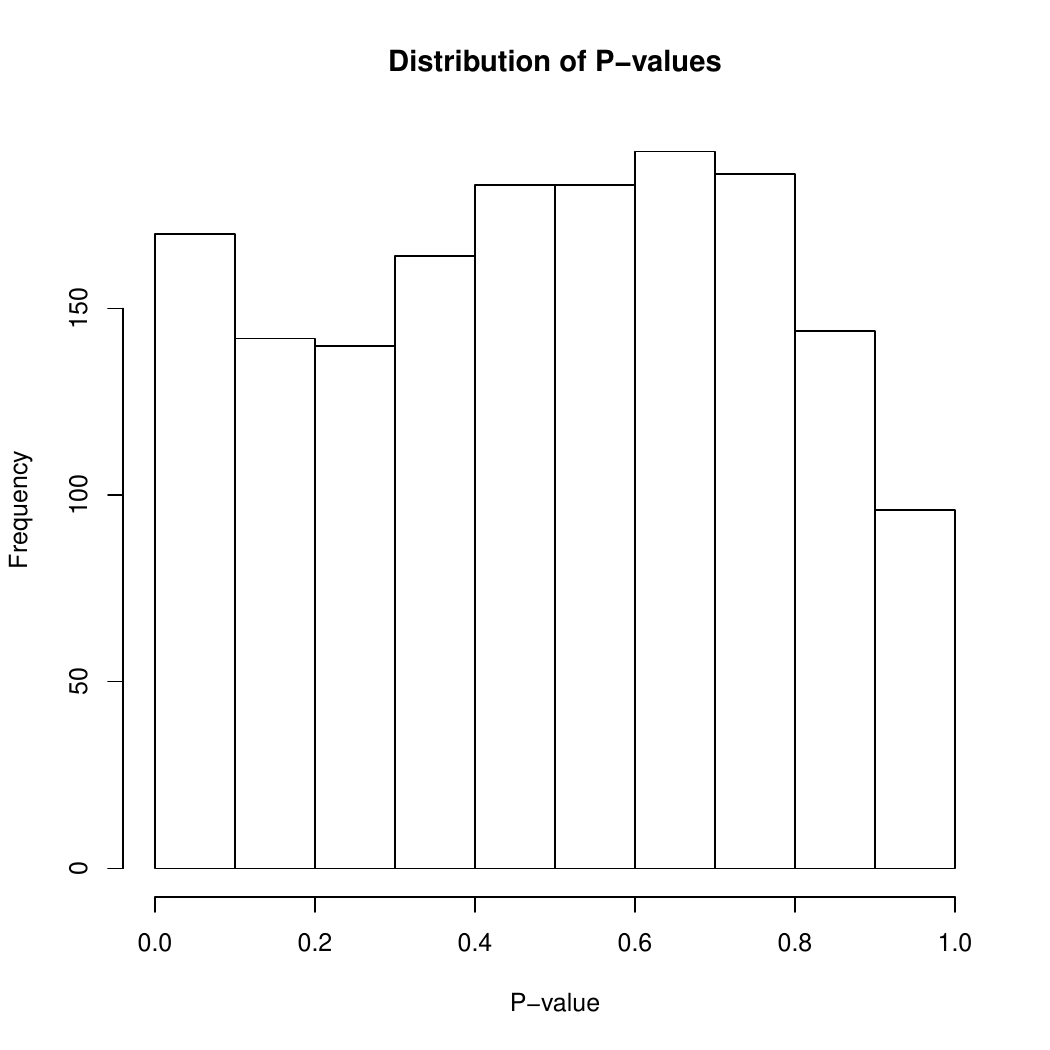} 
	\includegraphics[scale=0.3]{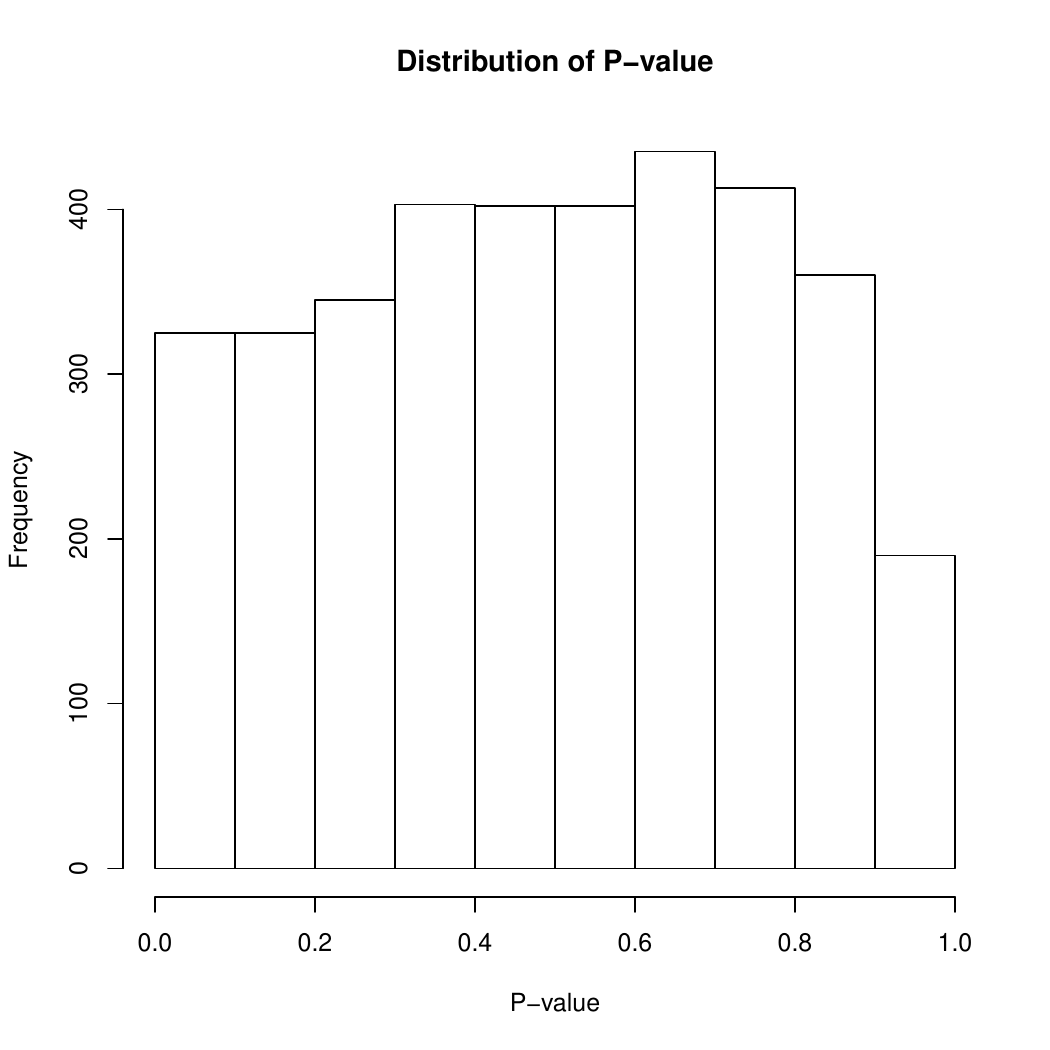} 
	\caption{Distribution of p-values obtained from the null $F$-distribution for FSR for DMN4 (left) and DMN6 (right).}
		\label{fig:figure1FSR}
	\end{figure}

\pagebreak

%\begin{figure}[h]
%\centering
%\includegraphics[scale=0.45]{DMN4_FSR.pdf} 
%%\hspace{-2.75em} 
%\includegraphics[scale=0.45]{DMN6_FSR.pdf} 
%\caption{Distribution of p-values obtained from the $F$ null distribution for FSR for DMN4 (left) and DMN6 (right).}
%\end{figure}

%
%
%\begin{figure}[h]
%\centering
%\begin{tabular}{cc}
%\includegraphics[scale=0.3]{../Figures/DMN4_FSR.pdf} &
%%\hspace{-2.75em} 
%\includegraphics[scale=0.3]{../Figures/DMN6_FSR.pdf} 
%\end{tabular}
%\caption{Distribution of p-values obtained from the $F$ null distribution for FSR for DMN4 (left) and DMN6 (right).}
%\end{figure}

%\begin{figure}[h]
%\centering
%%\begin{tabular}{cc}
%\hspace{-2.8em}\includegraphics[scale=0.6]{/Users/farouknathoo/Desktop/DCM_Genetics/figures/LASSO_PATH_genetic} 
%\hspace{-2.8em}
%%\end{tabular}
%\caption{The regularization path obtained from symmetric multinomial logistic regression with LASSO penalty with the priority set of 100 SNPs identified above the cut-off in Figure 3 included in the model.} 
%\label{fig:figure3}
%\end{figure}

%\begin{figure}[h]
%\centering
%%\begin{tabular}{cc}
%\includegraphics[scale=0.6]{/Users/farouknathoo/Desktop/DCM_Genetics/figures/LASSO_PLOT_full} 
%\hspace{-2.8em}
%%\end{tabular}
%\caption{The regularization path obtained from symmetric multinomial logistic regression with LASSO penalty with the priority set of 100 SNPs identified above the cut-off in Figure 3 included in the model along with non-genetic covariates representing age, sex, handedness, and education.} 
%\label{fig:figure3}
%\end{figure}

\pagebreak

\section*{Tables}
%

%
%\section*{Tables}
%

%

%\begin{table}[h]
%\centering
%\begin{tabular}{ll}
%  \hline
% Abbreviation & Description \\ 
%  \hline
%  APOE & Apolipoprotein E\\
%  BOLD &  Blood-Oxygen-Level Dependent
%  DMN & Default Mode Network \\
%rs-fMRI & resting-state functional magnetic resonance imaging\\
%ADNI & Alzheimer\'s Disease Neuroimaging Initiative\\
%DCM & dynamic causal model\\
%DTI & diffusion tensor imaging (DTI)\\
%TE & echo time\\
%SNP & single nucleotide polymorphisms\\
%LME & linear mixed effect\\
%FSL & a comprehensive library of analysis tools for FMRI, MRI and DTI brain imaging data
%FSR & function on scalar regression\\
%AD & Alzheimer\'s Disease\\
%MMSE & Mini Mental State Exam\\
%MNI & Montreal Neurological Institute\\
%MPRAGE &  magnetization-prepared rapid gradient echo\\
%NINCDS-ADRA & Neurological and Communicative Disorders and Stroke and the Alzheimer?s Disease Related Disorders Association
%SPM12 & statistical parametric mapping 12\\
%ICA & independent component analysis\\
%GWAS & genome-wide association study\\
%PACE & Principal Analysis by Conditional Expectation\\
%PLINK & 
%PET & positron emission tomography\\
%WMS & Wechsler Memory Scale\\
%PLINK & whole genome association analysis toolset
%  \hline
%\end{tabular}
%\caption {Table of abbreviations for the manuscript.} \label{tab:abbrev} 
%\end{table}

\begin{table}[h]
\centering
\begin{tabular}{r|llrr}
  \hline
 & DMN Region & MNI Coordinate \\ 
  \hline
1 & Posterior cingulate (PCC)& 0, -52, 26 \\ 
  2 & Medial Prefrontal (MPFC) & 3, 54, -2 \\ 
  3 & Left intraparietal cortex (LIPC) & -50, -63, 32  \\ 
  4 & Right intraparietal cortex (RIPC) & 48, -69, 35\\
  5 & Left inferior temporal (LIT) & -61, -24, -9 \\ 
  6 & Right inferior temporal (RIT) & 58, -24, -9  \\ 
%  7 & Left Hippocampus (LH) & -29, -19, -15  \\ 
%  8 & Right Hippocampus (RH) & 25, -16, -20 \\ 
  \hline
\end{tabular}
\caption {The MNI coordinates associated with the regions of interest in our study. The first four rows correspond to the network DMN4 while DMN6 corresponds to all eight rows.} \label{tab:DMN6H_region} 
\end{table}

\pagebreak

\begin{table}[htp]

\caption{Distribution of demographic variables (obtained at the baseline visit) across disease groups within our sample of 111 subjects. The p-values in the final column are based on a one-way ANOVA for continuous variables and a Fisher's exact test for categorical variables.}
\begin{center}
\begin{tabular}{l l l l l l}
\hline
& & AD & MCI & CN & p-value\\
\hline 
n & & 12 & 63 & 36 & \\
\hline
GENDER (\% of group) & Female & 8 (66.7) & 34 (54.0)& 23 (63.9) & 0.6402\\ 

& Male & 4 (33.3) & 29 (46.0)& 13 (36.1) & \\ 
\hline
HAND (\% of group) & Left & 1 ( 8.3) & 2 ( 3.2) & 2 ( 5.6) & 0.5036 \\ 
& Right & 11 (91.7) & 61 (96.8)& 34 (94.4)&\\ 
\hline 
Age - mean (sd) & & 75.82 (7.91)& 72.70 (7.66)& 75.35 (6.73)& 0.15\\ 
\hline
EDUCATION - mean (sd)& & 16.33 (2.53)& 16.06 (2.66)& 16.19 (2.14)& 0.929 \\ 
\hline
APOE $\epsilon$4 Alleles (\% of group)& Zero& 1 ( 8.3)& 35 (55.6)& 24 (66.6)& 0.0045\\ 
& One & 9 (75.0) & 22 (34.9) & 11 (30.6) & \\
& Two & 2 (16.7) & 6 ( 9.5) & 1 ( 2.8) & \\
\hline 
\end{tabular}
\end{center}
\label{default}
\end{table}
\pagebreak

\begin{table}[]
\centering
\begin{tabular}{c|llllrc}
 \hline
Rank & Network Edge & SNP & p-value & B-p-value	& q-value  &	Chromosome \\ 
  \hline
 1 & LIPC $\to$ PCC & kgp9433690\_G & 0.00025 &0.00040 & 0.28  & 9 \\
 2 &  LIPC $\to$ PCC & rs13287994\_A & 0.00035 &0.00065 & 0.28 & 9\\
3  &  PCC $\to$ PCC  & rs9317920\_G & 0.0013 &0.0022 & 0.45 & 13 \\
4 & LIPC $\to$ RIPC & rs13287994\_A & 0.0020&0.0026 & 0.45  & 9 \\
5  & LIPC $\to$ PCC & kgp4931190\_C & 0.0021&0.0030 & 0.45  & 9 \\
6  &  PCC $\to$ PCC & kgp12216228\_G &0.0025 &0.0036 & 0.45  & 13 \\
7  &  PCC $\to$ PCC &  rs1935110\_T & 0.0025&0.0033 & 0.45 & 13 \\
8  & PCC $\to$ MPFC & kgp1147116\_A & 0.0027&0.0039 & 0.45 & 9 \\
9 & LIPC $\to$ RIPC & kgp4931190\_C & 0.0028&0.0042 & 0.45 & 9 \\
10 & LIPC $\to$ LIPC &  rs2646852\_G & 0.0030&0.0033 & 0.45 & 1\\
%11 & LIPC $\to$ RIPC & kgp9433690\_G & 0.0030597466&0.003749813 & 0.4450541 & 9 \\
%12 &  PCC $\to$ PCC &  rs7335304\_A & 0.0043345291&0.006749663 & 0.5033231 & 13 \\
%13 & MPFC $\to$ LIPC & kgp320179\_A & 0.0046746950&0.004949753 & 0.5033231 & 18 \\
%14 & MPFC $\to$ MPFC &  rs2646852\_G & 0.0047142735&0.00655 & 0.5033231  & 1 \\
%15 & PCC $\to$ PCC & kgp5238984\_T & 0.0050448144 &0.006899655 & 0.5033231  & 3 \\
%16 &  PCC $\to$ MPFC & rs1625700\_A & 0.0053308976&0.00779961 & 0.5033231 & 9   \\ 
%17 & LIPC $\to$ MPFC & kgp5238984\_T & 0.0054419745&0.00605  & 0.5033231  & 3   \\
%18 &  PCC $\to$ MPFC & kgp8158375\_C & 0.0062065704&0.00779961 &0.5033231  & 9 \\
%19 & RIPC $\to$ RIPC & rs9317920\_G & 0.0070203068&0.00905 & 0.5033231   & 13  \\
%20 & RIPC $\to$ LIPC & rs10816805\_T & 0.0070766891&0.0099 & 0.5033231 &9   \\
  \hline
\end{tabular}
\caption {results of the linear mixed effects model longitudinal analysis of pairs of network edges with SNPs that are relatively highly ranked based on the p-values, parametric-bootstrap p-values (B-p-value) and q-values (after adjusting for 1600 tests) performed using the DMN4 PACE data.
} \label{tab:DMN4_LME} 
\end{table}

\begin{table}[]
\centering
\begin{tabular}{c|llllrc}
\hline 
Rank & Network Edge & SNP & p-value	&  B-p-value  &q-value&	Chromosome  \\  \hline
1 & LIPC$\to$ RIPC & rs13287994\_A & 0.00074 &0.0010& 0.24 &9 \\ 
  2 & RIPC$\to$ PCC & rs7935380\_T & 0.00093 &0.0010& 0.24 & 11\\ 
  3 & RIPC$\to$ PCC & rs2646852\_G & 0.0011 &0.0014& 0.24&1 \\ 
  4 & MPFC$\to$ MPFC & kgp2936399\_A & 0.0012 &0.0013& 0.24 &5 \\ 
  5 & LIPC$\to$ RIPC & kgp9433690\_G & 0.0013 &0.0022& 0.24&9 \\ 
  6 & LIPC $\to$ RIPC & kgp4931190\_C & 0.0014 &0.0014& 0.24& 9\\ 
  7 & RIPC $\to$ RIPC & kgp12216228\_G & 0.0014 &0.0020& 0.24&13 \\ 
  8 & RIPC $\to$ RIPC & rs1935110\_T & 0.0014 &0.0020& 0.24&13 \\ 
  9 & MPFC $\to$ MPFC & rs2646852\_G & 0.0014 &0.0023& 0.24&1 \\ 
  10 & RIPC $\to$ RIPC & rs9317920\_G & 0.0017 &0.0029& 0.24&13 \\ 
%  11 & LIPC $\to$ MPFC & kgp6435439\_G & 0.0017995 && 0.2432141&3 \\ 
%  12 & RIPC $\to$ RIPC & rs7335304\_A & 0.0018241 && 0.2432141& 13\\ 
%  13 & LIPC $\to$ MPFC & kgp9433690\_G & 0.0022594 && 0.2688732& 9\\ 
%  14 & LIPC $\to$  PCC & kgp9433690\_G & 0.0025162 && 0.2688732&9 \\ 
%  15 & RIPC $\to$ PCC & rs11601321\_G & 0.0025207 && 0.2688732&11 \\ 
%   16 & PCC $\to$ PCC & rs9317920\_G & 0.0027850 && 0.2784954&13 \\ 
%  17 & LIPC $\to$ MPFC & kgp4931190\_C & 0.0030688 && 0.2888281&9 \\ 
%  18 & LIPC $\to$ PCC & rs13287994\_A & 0.0038676 && 0.2953928&9 \\ 
%  19 & PCC $\to$ PCC & kgp12216228\_G & 0.0040843 && 0.2953928&13 \\ 
%  20 & PCC $\to$ PCC & rs1935110\_T & 0.0040843 && 0.2953928 &13\\ 
  \hline
\end{tabular}
\caption {results of the function-on-scalar regression longitudinal analysis of pairs of network edges with SNPs that are relatively highly ranked based on the  p-values and q-values after adjusting for 1600 tests performed using the DMN4 data.} \label{tab:DMN4} 
\end{table}

\begin{table}[]
\centering
\begin{tabular}{ccllc}
\hline 
Rank LME & Rank FSR & Network Edge & SNP & Chromosome \\  \hline
4 & 1 &LIPC $\to$ RIPC & rs13287994\_A& 9 \\
9 &6 & LIPC $\to$ RIPC & kgp4931190\_C& 9\\
11 &5 & LIPC $\to$ RIPC & kgp9433690\_G& 9 \\
1 & 14 & LIPC $\to$ PCC & kgp9433690\_G& 9 \\
2 &  18& LIPC $\to$ PCC & rs13287994\_A& 9\\
3  & 16 &PCC $\to$ PCC  & rs9317920\_G& 13\\
6  & 19&  PCC $\to$ PCC & kgp12216228\_G& 13 \\
7 &20  &  PCC $\to$ PCC &  rs1935110\_T& 13\\
  \hline
\end{tabular}
\caption {the rank (out of 1600) of the intersection of the top 20 DMN4 connection-SNP pairs for FSR and LME combined. These associations are highlighted as potential signals as a result of their stability.} \label{tab:DMN4_Intersect} 
\end{table}

\begin{table}[]
\centering
\begin{tabular}{c|lllllc}
 \hline
Rank & Network Edge & SNP & p-value & B-p-value & q-value &	Chromosome  \\ 
  \hline
1 &  LIPC $\to$ PCC & rs13287994\_A &7.33e-05&2.00e-04 & 0.18 & 9\\
2 &  LIPC $\to$ PCC & kgp9433690\_G & 1.21e-04&3.00e-05 & 0.18  & 9 \\
3 &  LIPC $\to$ LIT & kgp9433690\_G &1.59e-04& 9.99e-05 & 0.18 & 9 \\
4 &  LIPC $\to$ LIT & rs13287994\_A &2.041e-04&1.50e-04  & 0.18  & 9 \\
5 & MPFC $\to$ MPFC & rs17102906\_C & 3.61e-04&1.50e-04 & 0.26  & 11 \\
6 &  RIPC $\to$ RIT & kgp10801842\_G & 4.52e-04&3.50e-04 & 0.27  & 5 \\
7 &  LIPC $\to$ LIT & kgp4931190\_C &6.25e-04&4.50e-04 & 0.29  & 9 \\
8 &  LIT $\to$ LIT & kgp3071374\_C &7.53e-04&4.70e-04 & 0.29  & 6  \\
9 &  LIT $\to$ LIT &  rs9388153\_T &8.63e-04&5.20e-04  & 0.29 & 6  \\
10 & LIPC $\to$ PCC & kgp4931190\_C &8.69e-04&0.0012 & 0.29  & 9  \\
%11& MPFC $\to$ MPFC & kgp3071374\_C &8.794008e-04&0.001949903 & 0.2878039  & 6   \\
%12 & LIT $\to$ LIPC &  rs6021246\_C &1.044492e-03&0.001050053 & 0.3133477 & 20  \\
%13 & RIPC $\to$ RIT &  rs1417416\_A &1.177931e-03&0.001249938 & 0.3261962 & 1 \\
%14 & MPFC $\to$ MPFC &  rs9388153\_T &1.432845e-03&0.00135 & 0.3441976 & 6  \\
%15 & LIPC $\to$ RIPC & rs13287994\_A &1.434157e-03&0.001849908 & 0.3441976  &9   \\
%16 & LIPC $\to$ LIPC & rs13287994\_A &2.169975e-03&0.003299835 & 0.4882443  & 9  \\
%17 &  LIT $\to$ LIT & rs17102906\_C &2.392706e-03&0.003349833 & 0.5066908 &  11 \\
%18 &  RIT $\to$ RIT & rs17102906\_C &2.861076e-03&0.004299785 & 0.5722152 &11   \\
%19 & PCC $\to$ RIPC &   rs949200\_T &3.383106e-03&0.004949753 & 0.5848051 & 18 \\
%20 & PCC $\to$ RIPC & kgp8588069\_A &3.383106e-03&0.004849758 & 0.5848051 & 18 \\
\hline
\end{tabular}
\caption {the results of the linear mixed effects model longitudinal analysis of pairs of network edges with SNPs that are relatively highly ranked based on the p-values, parametric-bootstrap p-values (B-p-value) and q-values (after adjusting for 3600 tests) performed using the DMN6 PACE data.
} \label{tab:DMN8H_LME} 
\end{table}

\begin{table}[]
\centering
\begin{tabular}{c|lllllc}
\hline 
Rank & Network Edge & SNP & p-value & B-p-value	& q-value  &	Chromosome  \\  \hline
1 & RIPC $\to$ LIT & rs11601321\_G &2.00e-04 &1.00e-04& 0.30 &11 \\ 
  2 & LIT $\to$ LIT & rs17102906\_C & 2.70e-04 &2.00e-4& 0.30 &11 \\ 
  3 & RIT $\to$ RIT & rs17102906\_C & 3.50e-04  &7.00e-4& 0.30 &11 \\ 
  4 & LIT $\to$ LIT & kgp3071374\_C & 3.60e-4 &8.00e-4& 0.30 &6 \\ 
  5 & LIT $\to$ LIT & rs9388153\_T & 4.90e-04 &5.00e-4& 0.30 & 6\\ 
  6 & MPFC $\to$ MPFC & rs17102906\_C &5.10e-4  &7.00e-4& 0.30 & 11\\ 
  7 & RIPC $\to$ RIPC & kgp3071374\_C & 9.40e-4 &1.50e-3& 0.42 &6\\ 
  8 & MPFC $\to$ MPFC & kgp3071374\_C &1.07e-3  &1.40e-3& 0.42 &6\\ 
  9 & RIT $\to$ RIPC & rs3862175\_C &1.15e-3  &1.10e-3& 0.42 &18 \\
  10 & PCC $\to$ RIPC & rs949200\_T &1.28e-3 &1.40e-3& 0.42 &18 \\ 
  10 & PCC $\to$ RIPC & kgp8588069\_A &1.28e-3 &1.70e-3& 0.42 & 18\\ 
%  12 & RIPC $\to$ RIPC & rs9388153\_T & 0.0017360101 & 0.4859115 & 6\\ 
%  13 & RIT $\to$ RIPC & rs11873190\_T & 0.0018225877 & 0.4859115 & 18\\ 
%  14 & MPFC $\to$ MPFC & rs9388153\_T & 0.0018896557 & 0.4859115 & 6\\ 
%  15 & LIPC $\to$ PCC & rs4738020\_T & 0.0023346685 & 0.5603204 & 8 \\ 
%  16 & LIT $\to$ MPFC & kgp5238984\_T & 0.0027593323 & 0.5768492 &3 \\ 
%  17 & PCC $\to$ PCC & rs17102906\_C & 0.0028999843 & 0.5768492 &11\\ 
%  18 & RIT $\to$ PCC & rs4699458\_T & 0.0031167802 & 0.5768492 &18 \\ 
%  19 & LIT $\to$ LIPC & rs6021246\_C & 0.0032074765 & 0.5768492 &20 \\ 
%  20 & RIT $\to$ RIPC & kgp1838794\_A & 0.0033132866 & 0.5768492 &18 \\ 
  \hline
\end{tabular}
\caption {the results of the function-on-scalar regression longitudinal analysis of pairs of network edges with SNPs that are relatively highly ranked based on the  p-values, parametric-bootstrap p-values (B-p-value) and q-values after adjusting for 3600 tests performed using the DMN6 PACE data.} \label{tab:DMN6} 
\end{table}

\begin{table}[]
\centering
\begin{tabular}{ccllc}
\hline 
Rank LME & Rank FSR & Network Edge & SNP & Chromosome \\  \hline
5 &6& MPFC $\to$ MPFC & rs17102906\_C & 11 \\
14 & 14 & MPFC $\to$ MPFC &  rs9388153\_T & 6  \\
11& 8 &MPFC $\to$ MPFC & kgp3071374\_C & 6   \\
8 & 4& LIT $\to$ LIT & kgp3071374\_C & 6  \\
9 & 5& LIT $\to$ LIT &  rs9388153\_T & 6  \\
17 &2&  LIT $\to$ LIT & rs17102906\_C &  11 \\
18 &3 & RIT $\to$ RIT & rs17102906\_C &11   \\
12 & 19 & LIT $\to$ LIPC &  rs6021246\_C & 20  \\
19 & 10& PCC $\to$ RIPC &   rs949200\_T &18 \\
20 &10 & PCC $\to$ RIPC & kgp8588069\_A & 18 \\
  \hline
\end{tabular}
\caption {the rank (out of 3600) of the intersection of the top 20 DMN6 connection-SNP pairs for FSR and LME combined. These associations are highlighted as potential signals as a result of their stability.} \label{tab:DMN4_Intersect} 
\end{table}

\end{document}